\documentclass[aps,pra,10pt,twocolumn,nobalancelastpage]{revtex4-2}
\usepackage{graphicx}
\usepackage{amsmath}
\usepackage{amsfonts}
\usepackage{txfonts}
\usepackage{tikz} \usetikzlibrary{arrows.meta,patterns}
\usepackage{listings}
\usepackage{xcolor}
\usepackage{enumitem}
\usepackage{hyperref}

\hypersetup{colorlinks,linkcolor=blue,citecolor=blue,urlcolor=blue}
\definecolor{codegreen}{rgb}{0,0.43,0}
\definecolor{codegray}{rgb}{0.5,0.5,0.5}
\definecolor{codepurple}{rgb}{0.58,0,0.8}
\definecolor{codeblue}{rgb}{0,0,0.73}
\definecolor{backcolour}{rgb}{0.95,0.95,0.92}
\lstdefinestyle{mystyle}{
    backgroundcolor=\color{backcolour},   
    commentstyle=\color{codegreen},
    keywordstyle=\color{codeblue},
    numberstyle=\tiny\color{codegray},
    stringstyle=\color{codepurple},
    basicstyle=\ttfamily\footnotesize,
    breakatwhitespace=false,         
    breaklines=true,                 
    captionpos=b,                    
    keepspaces=true,                 
    numbersep=5pt,                  
    showspaces=false,                
    showstringspaces=false,
    showtabs=false,                  
    tabsize=2
}
\newcommand{\vect}[1]{\mathbf#1}
\newcommand{\nab}{\vect{\nabla}}
\newcommand{\keyw}[1]{\texttt{#1}}
\newcommand{\vin}{V_\mathrm{in}}
\newcommand{\vout}{V_\mathrm{out}}

\setlist[enumerate]{noitemsep}

\begin{document}

\title{RLC Parameters of a Two-Wire Line with the Finite Element Method}

\author{\rule{0pt}{12pt}{Marc Boul\'e}}
\affiliation{{\'Ecole de technologie sup\'erieure, Montr\'eal, Qu\'ebec, H3C~1K3, Canada}}
\date{\today}

\begin{abstract}
\begin{minipage}{5.3in}
This tutorial paper shows how to compute the DC (or low-frequency) resistance, inductance and capacitance of a pair of parallel wires using the finite element method. A three-dimensional infinite domain (open boundary) modeling of electrostatic and magnetostatic fields is presented, along with the electrokinetic formulation for the current flow inside the wires. The effects of the insulation and of a proposed physical defect in the wires are also considered. The open-source ONELAB software is used to perform the simulations and the code listing is provided. Comparisons using analytical models (when applicable) and the Altair Flux software are performed to help validate the simulations. 
\\ {\small This article appeared in \textit{Canadian Journal of Physics}, vol. 104: 1-11 (2026), and may be downloaded for personal use only.} 
{\footnotesize\url{https://doi.org/10.1139/cjp-2025-0170}}
\end{minipage}
\end{abstract}

\maketitle

\section{Introduction}
\label{sec_intro}

A pair of long cylindrical conductors has a resistance, an inductance and a capacitance (RLC), which in this work, are assumed to be per unit length. The capacitance is normally evaluated electrostatically, in which case no current is involved; conversely, the low-frequency inductance is typically evaluated magnetostatically, where steady currents are imposed and the skin effect is neglected. 
For the uninsulated wirepair (two infinite and parallel cylindrical conductors in vacuum), exact expressions for these parameters are known~\cite{NEFF1991}. For insulated conductors or more complex geometries, however, the analytical calculations quickly become infeasible and numerical simulations must be used. 

The focus of this paper is time-invariant electromagnetism, and its goal is to show how RLC values can be determined for a wirepair with insulated conductors, both with or without a physical defect (described in the next section). 
The use of second order elements and an exact modeling of the unbounded (infinite) domain are also presented. 
Being such a ubiquitous and fundamental electrical structure, the wirepair presents a compelling opportunity for developing a detailed case study. 
The finite element method (FEM) is used for the simulations, which must be three dimensional to support the defect. Unless more sophisticated coordinate transformations are employed~\cite{HAZIM2022}, 3D simulations are also typically required for twisted wires, for example.

Simulation models with planar or axis symmetries allow the use of 2D FEM applications such as FEMM~\cite{MEEKER2020} and Agros~\cite{KARBAN2025}, both of which are open-source and particularly well suited to electromagnetics in an academic setting~\cite{BOULE2014}. When 3D simulations are required, the tools involved inevitably become more elaborate, ranging from a suite of open-source applications such as Elmer FEM~\cite{ELMERWEB}, FreeFEM~\cite{FREEFEMWEB}, NGSolve~\cite{NGSOLVEWEB} and ONELAB~\cite{ONELABWEB}, to commercial solvers such as Altair Flux~\cite{ALTAIRWEB}, Ansys Maxwell~\cite{ANSYSWEB} and COMSOL Multiphysics~\cite{COMSOLWEB}. 

The general procedure for applying the finite element method can be summarized in the following steps~\cite{IOAN2021}: 
\begin{enumerate}
\item Create a CAD model and mesh it;
\item Declare boundary conditions, constraints and material properties;
\item Create the necessary function spaces for the unknown fields, and formulate their governing differential equations;
\item Generate the matrix representation of the discretized equations over the mesh and solve it;
\item Compute post-processing results such as auxiliary fields and integrals.
\end{enumerate}

Although the simulation codes shown in the following sections are specific to ONELAB (Open Numerical Engineering LABoratory), most of the concepts apply in other FEM applications. ONELAB is an open-source FEM software bundle comprising two distinct applications: Gmsh~\cite{GMSH2024}, a finite element mesh generator, and GetDP~\cite{GETDP2024}, a General environment for the treatment of Discrete Problems. These two applications can work independently, but provide many convenient features when used together, such as the ability to easily modify simulation parameters directly in the Gmsh user interface.

\section{Modeling the Problem}
\label{sec_prob}

The context for the simulation is a 22/2 AWG (American Wire Gauge) cable, consisting of two solid copper wires with polyethylene insulation around each conductor. 
The insulating jacket holding the wires together is omitted for the sake of simplicity.

\subsection{Geometrical Model}

\begin{figure}
\caption{Geometry and electromagnetic constants of the wirepair's conductors and insulators.\label{fig2dxy}}
\centerline{\input{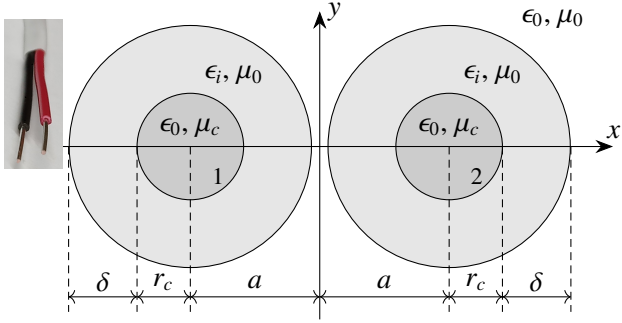}}
\end{figure} 

\begin{figure}
\caption{The wirepair segment to be analyzed (between $z=0$ and $z=\ell$).\label{fig2dxz}}
\centerline{\begin{tikzpicture}
\def\scl{1.2}

\def\iyt{1.514}
\def\iyb{0.05}
\filldraw [fill=gray!20!white, draw=gray!20!white] (-3.6,\iyb) rectangle (3.6,\iyt);
\draw[black] (-3.6,\iyb) -- (3.6,\iyb);
\draw[black] (-3.6,\iyt) -- (3.6,\iyt);
\filldraw [fill=gray!20!white, draw=gray!20!white] (-3.6,-\iyt) rectangle (3.6,-\iyb);
\draw[black] (-3.6,-\iyb) -- (3.6,-\iyb);
\draw[black] (-3.6,-\iyt) -- (3.6,-\iyt3);
\def\cyt{1.104}
\def\cyb{0.46}
\filldraw [fill=gray!40!white, draw=gray!40!white] (-3.6,\cyb) rectangle (3.6,\cyt) node[anchor=north east] {$2$};
\draw[black] (-3.6,\cyb) -- (3.6,\cyb);
\draw[black] (-3.6,\cyt) -- (3.6,\cyt);
\filldraw [fill=gray!40!white, draw=gray!40!white] (-3.6,-\cyt) rectangle (3.6,-\cyb) node[anchor=north east] {$1$};
\draw[black] (-3.6,-\cyb) -- (3.6,-\cyb);
\draw[black] (-3.6,-\cyt) -- (3.6,-\cyt3);
\draw[-{Stealth[scale=1.4]}] (-2.0,0) -- (-0.45,0) node[above = 1.5] {\scalebox{\scl}{$z$}};
\draw[-{Stealth[scale=1.4]}] (-2.0,0) -- (-2.0,2.0) node[right = 2] {\scalebox{\scl}{$x$}};
\draw[dashed,semithick] (-2.0,-2) -- (-2.0,2);
\draw[] (-1.5,-1.8) node {\scalebox{\scl}{$z=0$}};
\draw[dashed,semithick] (2.0,-2) -- (2.0,2);
\draw[] (2.5,-1.8) node {\scalebox{\scl}{$z=\ell$}};
\draw[->] (0.2,0.782) -- (0.6,0.782) node[right = 0.2] {\scalebox{\scl}{$I$}};
\draw[<-] (0.2,-0.782) -- (0.6,-0.782) node[right = -1.5] {\scalebox{\scl}{$I$}};
\coordinate (CTL) at (-2.0,0.782);
\coordinate (CTR) at (-2.0,-0.782);
\draw [stealth-stealth,thick,dashed] (CTL.west) to [out=-130,in=130] (CTR.west);
\draw[] (-2.65,0.24) node {\scalebox{\scl}{$\Delta V_0$}};
\draw[] (CTR) node[anchor=west] {\scalebox{\scl}{$-$}};
\draw[] (CTL) node[anchor=west] {\scalebox{\scl}{$+$}};
\draw[semithick] (-2.0,\cyb) -- (-2.0,\cyt);
\draw[semithick] (-2.0,-\cyb) -- (-2.0,-\cyt);
\coordinate (CTLL) at (2.0,0.782);
\coordinate (CTRL) at (2.0,-0.782);
\draw [stealth-stealth,thick,dashed] (CTLL.east) to [out=-50,in=50] (CTRL.east);
\draw[] (2.65,0.24) node {\scalebox{\scl}{$\Delta V_\ell$}};
\draw[] (CTRL) node[anchor=east] {\scalebox{\scl}{$-$}};
\draw[] (CTLL) node[anchor=east] {\scalebox{\scl}{$+$}};
\draw[semithick] (2.0,\cyb) -- (2.0,\cyt);
\draw[semithick] (2.0,-\cyb) -- (2.0,-\cyt);
\draw[] (-3,0.782) node[anchor=east] {\scalebox{\scl}{$\rho$}};
\draw[] (-3,-0.782) node[anchor=east] {\scalebox{\scl}{$\rho$}};
\draw[] (-3.6,0.782) node[anchor=east] {\scalebox{\scl}{\ldots}};
\draw[] (-3.6,-0.782) node[anchor=east] {\scalebox{\scl}{\ldots}};
\draw[] (3.6,0.782) node[anchor=west] {\scalebox{\scl}{\ldots}};
\draw[] (3.6,-0.782) node[anchor=west] {\scalebox{\scl}{\ldots}};
\end{tikzpicture}}
\end{figure}
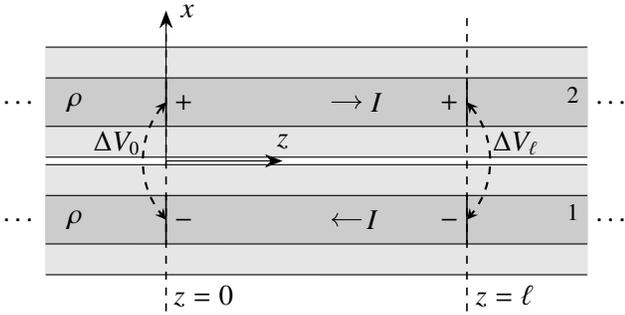

The dimensions of the wires are presented in Fig.~\ref{fig2dxy}, with $r_c=0.322$~mm, $\delta=0.41$~mm and $a=0.782$~mm. 
Figure~\ref{fig2dxz} shows the segment to be simulated, which is located between $z=0$ and $z=\ell$. 
Geometrical constants are specified in Fig.~\ref{figShared}, some of which must be visible to both GetDP and Gmsh. A length $\ell=4$~mm is simulated (called \keyw{ell} in the listing).
The model's code listing is specified in a \keyw{.geo} file, and is shown in Appendix~\ref{apx_geo}.

\begin{figure}
\caption{Geometrical constants  (\keyw{wirepaircommon.pro}).\label{figShared}}
\begin{lstlisting}[firstline=2, lastline=9]
// Geometrical constants (for Gmsh and GetDP)
mm  = 1E-3;         // units
ell = 4*mm;         // z length of segment
rc  = 0.322*mm;     // radius of conductor
ri  = rc + 0.41*mm; // outer radius of insulator
rb  = 2*mm;         // radius of inner boundary
re  = 2*rb;         // radius of external boundary
a   = ri + 0.05*mm; // x offset for center of wire
\end{lstlisting}
\end{figure}

\begin{figure}
\caption{Insulated wire with defect (top half of wire 2, partial 2D mesh).\label{figCut}}
\vspace{4pt}
\centerline{\includegraphics[width=3.35in]{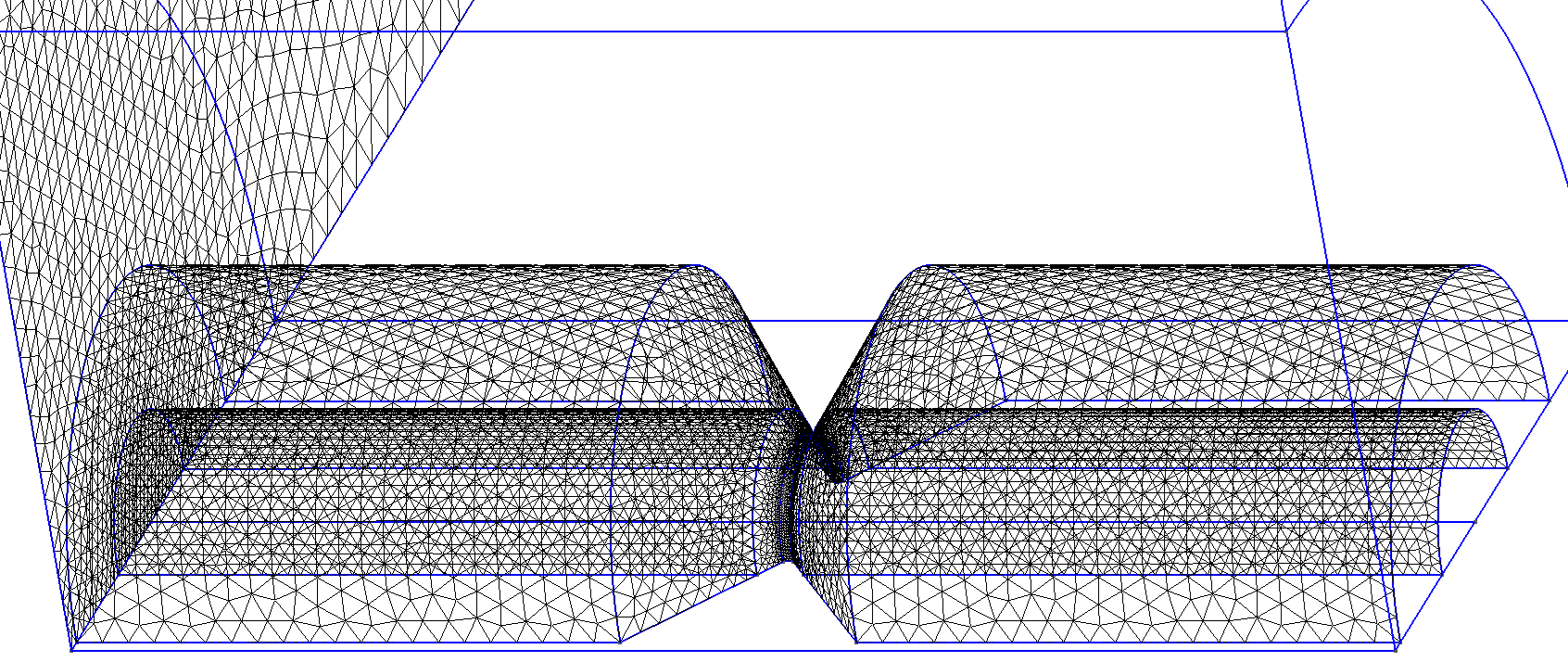}}
\end{figure}

An optional defect in the form of a ``V''-groove (not visible in Figs.~\ref{fig2dxy} and~\ref{fig2dxz}) is also modeled to evaluate its effect on the wirepair. 
This defect is shown in Fig.~\ref{figCut} and has the following characteristics: the ``V''-shaped cut is 0.1~mm deep into the conductor (before application of a \mbox{$20~\mu$m} fillet), and spans 0.14~mm in the longitudinal ($z$) direction at the conductor's surface. The cut extends outwards into the insulator and can be loosely imagined as resulting from the improper use of a wire stripper. 
The code listing with the defect is provided in the supplementary material~\cite{SUPMAT}. 

\subsection{Physical Model}
\label{ssec_phymod}

The conductors have a resistivity $\rho=1.7\cdot10^{-7}~\Omega$m, or equivalently, a conductivity $\sigma=1/\rho$. Since no dielectric polarization occurs within the conductors, they are assumed to have a permittivity of $\epsilon_0$. Their permeability is parametrized as a linear material for illustrative purposes; however, all simulations use $\mu_c=\mu_0$.  

The permeability of the insulation is assumed to be equal to $\mu_0$, and its conductivity is zero. The insulation is always present in the geometry description, and when its presence is not required, its permittivity is $\epsilon_i=\epsilon_0$ instead of $\epsilon_i=2.25\epsilon_0$. Vacuum surrounds the wires ($\epsilon_0$, $\mu_0$).

Given the symmetries involved, only one quadrant is modeled; this corresponds to the top half of wire~2 in Fig.~\ref{fig2dxy}. The left wire (wire~1) has a charge polarity and a current direction that are opposite to those of the right wire, hence the left wire can be replaced by a ground plane at $x=0$, consistent with the method of images~\cite{PAUL2008}. The potentials are also symmetrical about the $y=0$ plane. 
As simulated, the geometrical defect mentioned previously repeats every length~$\ell$. If only a single defect is desired, a longer length can be simulated without loss of generality, at the expense of added computation time and memory. The defect is also implicitly identical in both wires. 

Other simulation parameters appear in the top and bottom portions of Fig.~\ref{figPart1}, namely the voltage drop across the segment and other electrical properties. 
For wire~2, two potentials are defined according to Fig.~\ref{fig2dxz} such that $\vin=\Delta V_0/2$ and $\vout=\Delta V_\ell/2$. Currents can sometimes be imposed or automatically computed using global basis functions~\cite{DULAR1998}; however, these are not explored here. The resulting current is obtained instead from the power dissipated in the wires using a volume integral. 

\begin{figure}
\caption{Group and Function objects (\keyw{wirepair.pro}, part 1 of 7).\label{figPart1}}
\begin{lstlisting}[linerange={2-34}]
Include "wirepaircommon.pro";

// Constants
eps0  = 8.8541878188E-12;// ref permittivity (F/m)
mu0   = 1.25663706127E-6;// ref permeability (H/m)

// Simulation parameters
vIn   = 10.0;        // wire2 potential at z=0 (V)
vOut  = 9.9999;      // wire2 pot. at z=ell (V)
sigma = 1.0 / 1.7E-8;// conductivity (S/m)
epsRelInsul = 2.25;  // relative eps (unitless)
muRelCond   = 1.0;   // relative mu (unitless)

Group {
 VolCond   = #{100}; // conductor volume
 VolInsul  = #{101}; // insulator volume
 VolInt    = #{102}; // interior vacuum
 VolExt    = #{103}; // exterior vacuum (shell)
 SurLeft   = #{120}; // left boundary at x=0
 SurFront  = #{121}; // front plane at z=ell
 SurBack   = #{122}; // back plane at z=0
 SurExt    = #{123}; // curved ext. bound. at r=re
 VolNoCond = #{VolInsul, VolInt, VolExt};
 VolAll    = #{VolCond, VolNoCond};
 SurDiricA = #{SurLeft, SurExt, SurFront,SurBack};
}

Function {
 mu[VolCond]   = mu0 * muRelCond;
 mu[All]       = mu0; // all others
 eps[VolInsul] = eps0 * epsRelInsul; 
 eps[All]      = eps0; // all others
}
\end{lstlisting}
\end{figure} 

The following symbols are defined to simplify the presentation in the text, and are based on the group declarations appearing in the middle portion of Fig.~\ref{figPart1}:
\begin{equation}
\begin{split}
\Omega_c \equiv~\text{\keyw{VolCond};}\qquad &\Omega_n \equiv~\text{\keyw{VolNoCond};}\\ 	
\Omega=\Omega_c \cup \Omega_n & \equiv~\text{\keyw{VolAll}.}
\end{split}
\end{equation}

The three field regimes~\cite{IOAN2021,PIRIOU2024} involved in the RLC simulation are summarized next, where the order of the listed items is: the type of material property, the region involved, the unknown field to solve for, the type of differential equation used, the excitation source, the derived field(s).

Electrostatic: $\epsilon$, $\Omega_n$, scalar potential $V_n$, scalar Laplace, imposed potential, field $\vect{E}_n$.

Electrokinetic: $\sigma$, $\Omega_c$, scalar potential $V_c$, scalar Laplace, imposed potential, field $\vect{E}_c$ and current density $\vect{J}$.

Magnetostatic: $\mu$, $\Omega$, vector potential $\vect{A}$, vector Poisson, imposed current density, field $\vect{B}$.

Before presenting the mathematical models for the above regimes, a short description of boundary condition types is given next, along with the method used to handle the infinite domain.

\subsection{Boundary Conditions}

When an electric scalar potential is known on certain boundary surfaces, the values are imposed by what are known as Dirichlet (also called essential) boundary conditions. Electric fields are normal to Dirichlet surfaces with constant potentials. For magnetic vector potentials, surfaces of Dirichlet type typically imply tangent magnetic fields (and normal vector potentials), and do not usually involve imposed arbitrary vector potentials~\cite{ZHU2006}.

Conversely, when the electric field is tangent to certain surfaces of the domain, or when the magnetic field is perpendicular to certain surfaces, these can be imposed using a homogeneous Neumann (also called natural) boundary condition, for which the defining integral is equal to 0, as will be observed in the formulations in Sec.~\ref{ssec_mathmod}.

If the model has a repeating structure, such as the poles of an electric motor for example, the simulation domain can be simplified by treating only a portion of the larger geometry. This is implemented with a periodic boundary condition, which links the degrees of freedom between certain surfaces according to a given coefficient. More general terms like ``link'' or ``Dof map'' (Degree of freedom map) are sometimes used, instead of ``periodic''. For the simulation performed here, only Dirichlet and Neumann boundary conditions are used.

\subsection{Handling Infinite Domains}

\begin{figure}
\caption{Transformation of a point outside of the inner boundary (left) into the shell region (right). Locations of certain Dirichlet and Neumann boundary conditions are also shown.\label{figShell2D}}
\centerline{\begin{tikzpicture}
\def\scl{1.2}
\def\rb{2}
\def\rc{0.322}
\def\ri{0.732}
\fill[fill=gray!20!white] (-4,0) rectangle (-1,2.75);
\fill[fill=gray!20!white] (-4,2.75) .. controls (-2,3.2) and (-1.2,3.1) .. (-1,2.75);
\fill[fill=gray!20!white] (-1,2.75) .. controls (-0.6,2.05) and (-0.8,0.5) .. (-1,0);
\draw[fill=white      ] (-4,0) -- (-2,0) arc[start angle=0, end angle=90, radius=\rb] --cycle;
\draw[-{Stealth[scale=1.4]}] (-4,0) -- (-0.5,0) node[anchor=south] {\scalebox{\scl}{$x$}};
\draw[-{Stealth[scale=1.4]}] (-4.0,0) -- (-4,3.25) node[anchor=west] {\scalebox{\scl}{$y$}};
\draw[] (-4.25,1.3) node {\rotatebox{90}{Dirichlet}};
\draw[] (-2.5,-0.25) node {Neumann};
\draw[densely dashed] (-2.486,0) arc[start angle=0, end angle=180, radius=\ri];
\draw[densely dashed] (-2.896,0) arc[start angle=0, end angle=180, radius=\rc];
\coordinate (P) at (-2.5,2.25);
\fill[] (P) circle (0.06) node[anchor=west] {\scalebox{\scl}{\hspace{4pt}$p=(r,\theta,z)$}};
\draw[] (-3.3,1.45) node[] {\scalebox{\scl}{$r$}};
\draw[densely dotted] (-4,0) -- (P);
\draw[densely dotted] (-3.65,0) arc[start angle=0, end angle=60, radius=0.35];
\draw[] (-2.7,0.95) node[] {\scalebox{\scl}{$r_b$}};
\draw[densely dotted] (-4,0) -- ++(28:2);

\draw[fill=gray!20!white] (0.5,0) -- (3.3,0) arc[start angle=0, end angle=90, radius=2.8] --cycle;
\draw[fill=white] (0.5,0) -- (2.5,0) arc[start angle=0, end angle=90, radius=\rb] --cycle;
\draw[-{Stealth[scale=1.4]}] (0.5,0) -- (4.0,0) node[anchor=south] {\scalebox{\scl}{${x}$}};
\draw[-{Stealth[scale=1.4]}] (0.5,0) -- (0.5,3.25) node[anchor=west] {\scalebox{\scl}{${y}$}};
\draw[] (0.25,1.3) node {\rotatebox{90}{Dirichlet}};
\draw[] (2.0,-0.25) node {Neumann};
\draw[densely dashed] (2.014,0) arc[start angle=0, end angle=180, radius=\ri];
\draw[densely dashed] (1.604,0) arc[start angle=0, end angle=180, radius=\rc];
\coordinate (PT) at (1.725,1.837);
\fill[] (PT) circle (0.06) node[anchor=west] {\scalebox{\scl}{\hspace{4pt}$\tilde{p}=(\tilde{r},\theta,z)$}};
\draw[densely dotted] (0.5,0) -- (PT);
\draw[] (1.05,1.2) node[] {\scalebox{\scl}{$\tilde{r}$}};
\draw[densely dotted] (0.5,0) -- (PT);
\draw[densely dotted] (0.85,0) arc[start angle=0, end angle=60, radius=0.35];
\draw[] (1.8,0.95) node[] {\scalebox{\scl}{$r_b$}};
\draw[densely dotted] (0.5,0) -- ++(28:2);
\draw[] (2.7,0.60) node[] {\scalebox{\scl}{$r_e$}};
\draw[densely dotted] (0.5,0) -- ++(10:2.8);

\draw [->] (-2.4,2.4) to [bend left=40] (1.65, 1.95);


\end{tikzpicture}}
\end{figure}
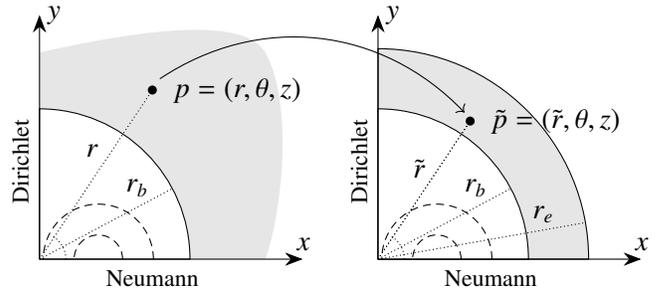

Modeling infinite domains can be done using many different methods in electrostatics and magnetostatics~\cite{CHEN1997}. The shell transformation is an exact method based on the function 
\begin{equation}
\label{eqShell}
\tilde{r} = r_e-\frac{r_b(r_e- r_b)}{r},
\end{equation}
where $r_b$ is the radius of the inner boundary and $r_e$ is the exterior radius of the shell. As shown in Fig.~\ref{figShell2D}, this transformation has the property of mapping all points $r > r_b$ outside of the inner boundary to points $r_b < \tilde{r} < r_e$ inside the shell region, thus mapping an infinite region to a finite one. 
The case $r_e=2r_b$ results in $d\tilde{r}/dr=1$ at $r=r_b$, thus providing a smooth connection at this boundary. The case $r_e = 0$ corresponds to the circle (or sphere) inversion, also called Kelvin transformation~\cite{MEEKER2020,NABIZADEH2021,REMACLE1994}, with its mapping $\tilde{r} = (r_b)^2/r$. 

The actual transformation required in the simulation is in the inverse direction of what is shown in eq.~(\ref{eqShell}) and Fig.~\ref{figShell2D} (and appears as the case $p=1$ in ref.~\cite{HENROTTE1999}), so that the finite elements in the shell region are deformed to cover the original infinite region. Such deformations are specified via their Jacobian matrices but do not need to be explicitly given in GetDP.
Given the spatial distortions involved, it is generally advisable to perform at least second order simulations when using the shell transformation (explained in Sec.~\ref{ssec_fs}).

The shell transformation can be applied to spherical or cylindrical boundaries in 3D. Given the wirepair's longitudinal structure, cylindrical coordinates $(r,\theta,z)$ can be used to illustrate the transformation (even if the simulation inherently uses Cartesian coordinates). If a field quantity needs to be evaluated during post processing at a point~$p$ that lies outside of the inner boundary, it must instead be evaluated at the point $\tilde{p}$ in the transformed (shell) region, according to Fig.~\ref{figShell2D} and eq.~(\ref{eqShell}).  

Another method for modeling infinite domains is the Improvised Asymptotic Boundary Condition (IABC)~\cite{MEEKER2014,MEEKER2013}. This method is based on a series of concentric thin shells with varying material properties, and can be used in practically any FEM software. Although not an exact method, it can still be very precise, as will be shown in Sec.~\ref{sec_res}. Obtaining correct field values for points that lie outside of the inner boundary is not possible; however, volume energy integrals are correctly evaluated on the entire domain.

\subsection{Mathematical Model}
\label{ssec_mathmod}

The differential equations of electrostatics (without free charges) and electrokinetics, in the regions $\Omega_n$ and $\Omega_c$ respectively, are the two scalar Laplace equations
\begin{equation}
\label{eqpoiV}
\nab\cdot\left(\epsilon\nab V_n\right)=0,\qquad \nab\cdot\left(\sigma\nab V_c\right)=0.
\end{equation}
The magnetostatic regime in region $\Omega$ is governed by the vector Poisson equation 
\begin{equation}
\label{eqpoiA}
\nab\times\left(\nab\times{\vect{A}}/{\mu}\right)=\vect{J}.
\end{equation}
The electric fields can be obtained with:
\begin{equation}
\label{eqmaxpot}
\vect{E}_n=-\nab V_n,\qquad \vect{E}_c=-\nab V_c ,
\end{equation}
while the current density and magnetic field are
\begin{equation}
\label{eqJE}
\vect{J}=\sigma\vect{E}_c=-\sigma\nabla V_c, \qquad   \vect{B}=\nab\times\vect{A}.
\end{equation}

Many finite element solvers employ a weighted residual method, of which the Galerkin method is a particular type: its defining characteristic is to use the same functions for both the basis and weighting functions. Each differential equation must also be given in what is known as a weak formulation~\cite{PIRIOU2024,ZHU2006,JIN2014,BASTOS2014,BIRO1998}, whereby the solution exists in a larger function space. This is a type of Hilbert space ($H$) with constraints on the derivatives of the function. The solution function must also be square integrable (the field must have finite energy). 

For scalar potentials solved in weak form, gradient spaces $H(\mathrm{grad})$ are involved, while for the magnetic vector potential, $\vect{H}(\mathrm{curl})$ spaces are used. Weak solutions require multiplication by a weighting function of scalar or vectorial nature ($W$ or $\vect{W}$). These functions also exist in similar function spaces, but with particular constraints on the Dirichlet boundaries ($\Gamma_D$), as will be shown below. Dirichlet and Neumann boundaries ($\Gamma_N$) form the entire boundary $\Gamma$ of a given volume $\Omega$, and $\vect{n}$ defines the local unit vector normal to $\Gamma$ (pointing outwards).

In the nonconducting region ($\Omega_n$), the weak electrostatic potential formulation is
\begin{equation}
\label{eqfrmEsmN}
\int_{\Omega_n}{\epsilon\nab V_n\cdot\nab W_n  \,d\Omega_n}-
\int_{\Gamma_{nN}}{\epsilon\nab V_n\cdot\vect{n}\,W_n   \,d\Gamma_{nN}}
= 0,
\end{equation} 
where $V_n,W_n\in H(\mathrm{grad}, \Omega_n)$, $V_n$ is imposed on $\Gamma_{nD}$ and $W_n=0$ on $\Gamma_{nD}$. 

The Dirichlet surface ($\Gamma_{nD}$) is the $x=0$ plane and the lateral surface of the conductor, where the potentials are $V_n=0$ and $V_n=\vin$ respectively. The Neumann surface ($\Gamma_{nN}$) corresponds to the portions of the planes $y=0$ and $z=\{0,\ell\}$ that bound $\Omega_n$, along with the curved exterior boundary at $r=r_e$, where the electric field is tangent to all such surfaces. The dot product in the second integral above is therefore equal to zero and the term is simply omitted. In the nonconducting region, the weak electrostatic potential formulation becomes
\begin{equation}
\label{eqfrmEsm}
\int_{\Omega_n}{\epsilon\nab V_n\cdot\nab W_n  \,d\Omega_n} = 0.
\end{equation} 

In a similar manner, the weak electrokinetic formulation in the conducting region (without the surface integral term) is
\begin{equation}
\label{eqfrmEkm}
\int_{\Omega_c}{\sigma\nab V_c\cdot\nab W_c  \,d\Omega_c} = 0,
\end{equation}
where $V_c,W_c\in H(\mathrm{grad}, \Omega_c)$, $V_c$ is imposed on $\Gamma_{cD}$ and $W_c=0$ on $\Gamma_{cD}$. The front and back faces of the conductor at $z=\{0,\ell\}$ constitute the Dirichlet surface ($\Gamma_{cD}$), where the potentials are respectively $V_c=\{\vin,\vout\}$. All other surfaces bounding the conductor in $\Omega_c$ are part of the Neumann surface ($\Gamma_{cN}$), where the electric field is tangent.

Using the substitution for $\vect{J}$ in eq.~(\ref{eqJE}), the weak magnetostatic formulation of eq.~(\ref{eqpoiA}) is 
\begin{equation}
\label{eqfrmBsm}
\int_\Omega{\left(\frac{\nab\times\vect{A}}{\mu}\right)\cdot\left(\nab\times\vect{W}\right)  d\Omega}\\
+\int_{\Omega_c}{\left(\sigma\nabla V_c\right)\cdot\vect{W}\, d\Omega_c} = 0,
\end{equation}
where $\vect{A},\vect{W}\in \vect{H}(\mathrm{curl}, \Omega)$ and $\vect{n}\times\vect{A}=\vect{n}\times\vect{W}=\vect{0}$ on $\Gamma_{D}$.  
The Neumann surface is the $y=0$ plane, where the magnetic field is normal to this surface, while all other exterior surfaces of $\Omega$ are of Dirichlet type with a tangent magnetic field. The surface integral term implicit in eq.~(\ref{eqfrmBsm}) is 
\begin{equation}
\label{eqresB}
\int_{\Gamma_N}{\left(\vect{n}\times\left(\nab\times{\vect{A}}/{\mu}\right)\right)\cdot\vect{W}\,  d\Gamma_N},
\end{equation}
but is not required since the cross product with $\vect{n}$ is equal to zero. 

Equations (\ref{eqfrmEsm}-\ref{eqfrmBsm}) are the equations that need to be solved in the numerical computation. In general, $\epsilon$, $\mu$ and $\sigma$ must not systematically be factored out of the integrals since they are not necessarily constant. The region dependent symbols \keyw{eps[]} and \keyw{mu[]} are two examples of this.

\subsection{RLC Calculations}

Once the potentials and their derived fields are obtained, the RLC calculations can be performed based on power and energy integrals.  
Computing the dissipated power ($P_c$) in the conducting region is done with the integral
\begin{equation}
\label{eqpc}
P_c=\int_{\Omega_c}\sigma \|\vect{E}_c\|^2\,d\Omega_c,
\end{equation}
while the energies in the magnetic and electric fields, $W_B$ and $W_E$, can be obtained with the integrals
\begin{equation}
\label{eqwbwe}
W_B=\int_{\Omega}\frac{1}{2\mu} \|\vect{B}\|^2\,d\Omega,
\qquad
W_E=\int_{\Omega_n}\frac{\epsilon}{2} \|\vect{E}_n\|^2\,d\Omega_n,
\end{equation}
where $\mu$ and $\epsilon$ represent the piecewise permeability and permittivity, as defined in the bottom part of Fig.~\ref{figPart1}.

Once these three integrals are evaluated in their appropriate regions, the resistance, inductance and capacitance can then be calculated by using the formulas below. 
The resistance per unit length ($R$) and the current ($I$) of the wirepair are
\begin{equation}
\label{eqresIR}
R=\frac{2\left(\vin-\vout\right)}{\ell I},\qquad I=\frac{P_c}{2\left(\vin-\vout\right)},
\end{equation}
while the inductance per unit length ($L$) and capacitance per unit length ($C$) are obtained by using
\begin{equation}
\label{eqresLC}
W_B=\frac{1}{2}L\ell I^2,\qquad 
W_E=\frac{1}{2}C\ell\left(2\vin\right)^2.
\end{equation}
All quantities involving the current should only be calculated when $\vin\neq\vout$. On the contrary, the capacitance above is only relevant when $\vin=\vout$.

\section{Simulation Setup}
\label{sec_sim}

The simulation is specified using ten types of objects in GetDP: Group, Function, Constraint, FunctionSpace, Jacobian, Integration, Formulation, Resolution, PostProcessing and PostOperation, each of which is explained next.

\subsection{Group and Function}
\label{ssec_group}

The first object to specify is the Group object (middle part of Fig.~\ref{figPart1}), where different entities such as surfaces and volumes are declared. Primary entities defined with numbers relate to the physical entities appearing in the geometry file (Appendix~\ref{apx_geo}).

Since the $y=0$ plane is a Neumann-type boundary for all potentials, it is not involved in any surface defined in the Group object. The Function object, also shown in Fig.~\ref{figPart1}, is useful for declaring the piecewise material properties mentioned above (permittivity and permeability) in the different regions.

\subsection{Constraint}
\label{ssec_cst}

The role of the Constraint object in Fig.~\ref{figPart2} is to specify the different boundary conditions for the fields to be solved. The constraints acting on the electric potentials inside and outside of the conductor are \keyw{CstVc} and \keyw{CstVn} respectively, while the constraint for the magnetic vector potential is \keyw{CstA}. 

\begin{figure}
\caption{Constraint object (\keyw{wirepair.pro}, part 2 of 7).\label{figPart2}}
\begin{lstlisting}[firstline=36, lastline=51]
Constraint {
 { Name CstVn; Case {
  { Region SurLeft; Value 0; }
  { Region VolCond; Value vIn; }
 }} // Neumann Vn: y=0, front, back, ext
 { Name CstVc; Case {
  { Region SurBack; Value vIn; }
  { Region SurFront; Value vOut; }
 }} // Neumann Vc: lateral conductor face
 { Name CstA;  Case {
  { Region SurDiricA; Value 0; }
 }} // Neumann A: y=0
 { Name CstGaugeA; Case { 
  { Region VolAll; SubRegion SurDiricA; Value 0; }
 }}
}
\end{lstlisting}
\end{figure}

The two main types of constraints are homogeneous Neumann (which do not require explicit declarations) and Dirichlet. Although it is possible to impose periodic boundary conditions (\keyw{Link} keyword) connecting the front and back planes, Neumann and Dirichlet boundary conditions are sufficient given the symmetries involved (and computationally more efficient for the magnetic vector potential, since slightly fewer degrees of freedom are required). 

In principle, the lateral surface of the conductor could be used in both the \keyw{CstVc} (in $\Omega_c$) and \keyw{CstVn} (in $\Omega_n$) constraints, but is not needed since this surface is a Neumann-type boundary in \keyw{CstVc}, while in \keyw{CstVn}, the potential \keyw{vIn} can simply be assigned to all of the nodes in \keyw{VolCond} (which implicitly includes the surface nodes). Another simplification that has no impact on the simulation time consists in using the entire front and back planes in \keyw{CstVc}, even though the majority of the surface area of these planes is outside of $\Omega_c$.

A gauging constraint (\keyw{CstGaugeA}) is also used for the magnetic vector potential. Although it is not formally required when $\vect{A}$ is used to obtain the magnetic field~\cite{JIN2014}, gauging can reduce the number of unknowns to solve for, and it can help the numerical procedure to converge more rapidly by improving numerical stability during resolution.

Gauging is typically implemented by imposing $\nabla\cdot\vect{A}=0$ (known as the Coulomb gauge) in the problem's formulation, or by forcing certain degrees of freedom to~0 on the edges of a spanning tree~\cite{CREUSE2019} (known as the tree/co-tree gauge). The latter method is used in this work. 
The \keyw{SubRegion} for the gauge constraint in Fig.~\ref{figPart2} is the \keyw{SurDiricA} surface defined in Fig.~\ref{figPart1}. This subregion corresponds to all surfaces where the potential is constrained, namely only Dirichlet surfaces in this case; certain surfaces involved in periodic links must also be included but are not applicable here.

\subsection{Function Space}
\label{ssec_fs}

In GetDP, scalar fields suitable for potentials are of type \keyw{Form0}. This is the type of space required for the electric potential and is used in the first two function spaces in Fig.~\ref{figPart3}.  
A curl-conforming function space is instead used for the magnetic vector potential. This can be specified using the type \keyw{Form1}, and is the third function space declared. 

\begin{figure}
\caption{FunctionSpace object (\keyw{wirepair.pro}, part 3 of 7).\label{figPart3}}
\begin{lstlisting}[firstline=53, lastline=97]
FunctionSpace {
 { Name HgradVn; Type Form0; // V nonconductor
   BasisFunction {
    { Name snn; NameOfCoef vnn; Support VolNoCond;
      Function  BF_Node; Entity NodesOf[All];}
   }
   Constraint {
    { NameOfCoef vnn;  EntityType NodesOf; 
      NameOfConstraint CstVn; }
   }
 }
 { Name HgradVc; Type Form0; // V conductor
   BasisFunction {
    { Name snc; NameOfCoef vnc; Support VolCond; 
      Function  BF_Node; Entity NodesOf[All];}
   }
   Constraint {
    { NameOfCoef vnc;  EntityType NodesOf; 
      NameOfConstraint CstVc; }
   }
 }
 { Name HcurlA; Type Form1; // A all
   BasisFunction {
    { Name se;  NameOfCoef ae;  Support VolAll; 
      Function  BF_Edge; Entity EdgesOf[All];}
    { Name s3a; NameOfCoef a3a; Support VolAll;
      Function  BF_Edge_3F_a; 
      Entity    FacetsOf[All]; }
    { Name s3b; NameOfCoef a3b; Support VolAll;
      Function  BF_Edge_3F_b;   
      Entity    FacetsOf[All]; }
   }
   Constraint {
    { NameOfCoef ae;   EntityType EdgesOf;
      NameOfConstraint CstA; }
    { NameOfCoef a3a;  EntityType FacetsOf; 
      NameOfConstraint CstA; }
    { NameOfCoef a3b;  EntityType FacetsOf; 
      NameOfConstraint CstA; }
    { NameOfCoef ae;   EntityType EdgesOfTreeIn; 
      EntitySubType    StartingOn;
      NameOfConstraint CstGaugeA; }
   }
 }
}
\end{lstlisting}
\end{figure}

Derivatives of functions in function spaces can imply different differential operators. In a \keyw{Form0} space, the derivative can be identified as the gradient operator, and in a \keyw{Form1} space, the derivative can be identified as the curl operator, consistent with the function spaces given in Sec.~\ref{ssec_mathmod}. For this reason, using the more general exterior derivative symbol~\keyw{d} is recommended instead of the \keyw{Grad} and \keyw{Curl} keywords. 

Each function space's interpolation function is composed of one or more hierarchically structured basis functions~\cite{ZHU2006,GEUZAINE2001}. Each basis function has a type, a coefficient name, and a domain (support) on which it is defined; it is also where the previously declared constraints are applied. 
For \keyw{Form0} spaces, the first order basis functions are associated with the scalar field's values at the nodes of the elements (\keyw{{BF\char`_Node}}). For \keyw{Form1} spaces, the first order basis functions are associated with the components of the unknown vector field along the edges of the elements (\keyw{BF\char`_Edge}).

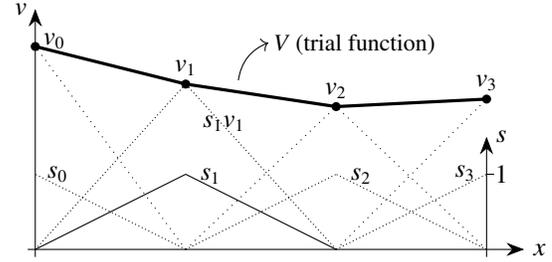
\begin{figure}
\caption{Triangle functions ($s_n$) and coefficients ($v_n$) for approximating a 1D scalar function over three line elements.\label{figShape}}
\centerline{\begin{tikzpicture}

\def\scl{1.1}

\def\xA{-3}
\def\xB{-1}
\def\xC{1}
\def\xD{3}

\def\yb{0}
\def\ys{1}
\def\yA{2.7}
\def\yB{2.2}
\def\yC{1.9}
\def\yD{2.0}

\draw[-{Stealth[scale=1.4]}] (-3.1,\yb) -- (3.5,\yb) node[anchor=west] {\scalebox{\scl}{$x$}};
\draw[-{Stealth[scale=1.4]}] (\xA,-0.1) -- (\xA,3.2) node[anchor=east] {\scalebox{\scl}{$v$}};
\draw[-{Stealth[scale=1.4]}] (\xD,-0.1) -- (\xD,1.5) node[anchor=west] {\scalebox{\scl}{$s$}};
\draw[] (\xD,\ys) -- (3.1,\ys);

\draw[densely dotted] (\xB,\yb) -- (\xA,\ys) node[anchor=west,shift={(1pt,0pt)}] {\scalebox{\scl}{$s_0$}};
\draw[] (\xA,\yb) -- (\xB,\ys) node[anchor=west,shift={(2pt,-0.5pt)}] {\scalebox{\scl}{$s_1$}};
\draw[] (\xC,\yb) -- (\xB,\ys);
\draw[densely dotted] (\xB,\yb) -- (\xC,\ys) node[anchor=west,shift={(2pt,-0.5pt)}] {\scalebox{\scl}{$s_2$}};
\draw[densely dotted] (\xD,\yb) -- (\xC,\ys);
\draw[densely dotted] (\xC,\yb) -- (\xD,\ys) node[anchor=east,shift={(-1pt,0.5pt)}] {\scalebox{\scl}{$s_3$}} node[anchor=west] {\scalebox{\scl}{$1$}};

\draw[dotted] (\xB,\yb) -- (\xA,\yA) node[anchor=west,shift={(0pt,2pt)}] {\scalebox{\scl}{$v_0$}};
\filldraw (\xA,\yA) circle (1.5pt);
\draw[densely dotted] (\xA,\yb) -- (\xB,\yB) node[anchor=south] {\scalebox{\scl}{$v_1$}};
\draw[densely dotted] (\xC,\yb) -- (\xB,\yB) node[anchor=west,shift={(30pt,15pt)}] {\scalebox{1.0}{$V$ (trial function)}};
\draw [->] (-0.3,2.25) to [bend left] (0.1,2.75);
\filldraw (\xB,\yB) circle (1.5pt) node[anchor=west,shift={(2.9pt,-15pt)}] {\scalebox{\scl}{$s_1v_1$}};
\draw[dotted] (\xB,\yb) -- (\xC,\yC) node[anchor=south] {\scalebox{\scl}{$v_2$}};
\draw[dotted] (\xD,\yb) -- (\xC,\yC);
\filldraw (\xC,\yC) circle (1.5pt);
\draw[dotted] (\xC,\yb) -- (\xD,\yD) node[anchor=south] {\scalebox{\scl}{$v_3$}};
\filldraw (\xD,\yD) circle (1.5pt);

\draw[line width=1.2pt] (\xA,\yA) -- (\xB,\yB) -- (\xC,\yC) -- (\xD,\yD) ;


\end{tikzpicture}}
\end{figure}

In 1D scalar problems, first order interpolation elements require only one type of basis function, composed of a set of triangle (hat) functions~\cite{BASTOS2014}. For the example shown in Fig.~\ref{figShape}, 
four functions ($s_n$) are required for the three line elements. These functions are multiplied by the coefficients ($v_n$) in order to represent the approximate solution $V$ (called trial function). 
In the example figure, the solution $V$ corresponds to $\sum_{n=0}^{3}s_nv_n$.

To improve the accuracy of the solution, a higher order interpolation can be obtained by adding higher order basis functions. For example, a set of basis functions that includes second order polynomials would allow for a much smoother solution in the example in Fig.~\ref{figShape}. 

Elements of second geometrical order (with curved edges) can also be used to improve simulation accuracy. Although geometrical and interpolation orders are two separate notions, they are used in tandem in this work, where both second order interpolation and geometrical elements are assumed. For geometries containing curved surfaces, it is generally better to use second order geometrical elements when they are available.

When second order geometrical elements are specified in the \keyw{.geo} file (last line in Fig.~\ref{figGeo}), the second order basis functions \keyw{BF\char`_Node\char`_2E} are automatically added to \keyw{Form0} spaces in GetDP. Second order basis functions must be explicitly added in \keyw{Form1} spaces. In a magnetic vector potential formulation, only non-gradient higher order functions need to be added. To achieve a second order interpolation space for \keyw{HcurlA}, the two facet functions \keyw{BF\char`_Edge\char`_3F\char`_a} and \keyw{BF\char`_Edge\char`_3F\char`_b} are included (Fig.~\ref{figPart3}). 
Since facet functions are not involved in the tree gauging, the \keyw{CstGaugeA} constraint is applied only to the single \keyw{BF\char`_Edge} function.

\subsection{Jacobian and Integration}
\label{ssec_ji}

The Jacobian object is used to apply geometric transformations to specified regions, suitable for handling infinite domains. The Jacobian for a cylindrical shell transformation~\cite{HENROTTE1999}, called \keyw{VolCylShell}, is used in the exterior region. 
The default Cartesian Jacobian, called \keyw{Vol}, is used in the main region. 
The \keyw{VolSphShell} Jacobian can be used to perform spherical shell transformations (or circular, in 2D planar cases) but is not applicable here. 
Jacobian specifications appear in the top half of Fig.~\ref{figPart4}. The two parameters of \keyw{VolCylShell} are the inner and outer radii of the shell.

Given the exact nature of the shell transformation, the inner boundary (of radius $r_b$) can be placed arbitrarily close to the wires, provided that the mesh is sufficiently refined in the (empty) shell region. The $r_b$ radius is nonetheless extended to 2~mm, to allow an undistorted view of the fields in the vicinity of the wires. 
The outer boundary's radius ($r_e$) is chosen to be exactly twice the inner boundary's radius, for a smooth transition of element distortions across the inner boundary~\cite{HENROTTE1999}.

\begin{figure}
\caption{Jacobian and Integration (\keyw{wirepair.pro}, part 4 of 7). \label{figPart4}}
\begin{lstlisting}[firstline=99, lastline=110]
Jacobian {
 { Name J1; Case { 
  { Region VolExt; Jacobian VolCylShell{rb,re}; }
  { Region All;    Jacobian Vol; } // all others
 }}      
}

Integration {
 { Name I1; Case {{ Type Gauss; Case { 
  { GeoElement Tetrahedron2; NumberOfPoints 4; }
 }}}}
}
\end{lstlisting}
\end{figure}

The Integration object in Fig.~\ref{figPart4} defines the interpolation points used for numerical integration in the elements (for Gaussian quadrature methods). 
For the regular formulations described previously, a minimal number of integration points based on the element orders is sufficient~\cite{GEUZAINE2001}. 
First order tetrahedral elements can be integrated with a single Gauss point (not used in this simulation), while four points are required for integrating in tetrahedra with second order basis functions~\cite{BASTOS2014,DHATT2012}.

\subsection{Formulation}
\label{ssec_formul}

Integrals in weak formulations (and in GetDP) correspond to inner products~\cite{JIN2014} of the form $\left\langle a,b\right\rangle$. 
In the weak formulations in eqs.~(\ref{eqfrmEsm}-\ref{eqfrmBsm}), the dot products are the comma separator in the inner product, and both sides of the dot product are the $a$ and $b$ quantities.  
In GetDP, a weighting function uses the same symbol as the unknown field ($V$ or $\vect{A}$ in this work). When the $V$ (or $\vect{A}$) symbol appears on the right side of the comma in the inner product, it implicitly represents the weighting function $W$ (or $\vect{W}$) that is associated with it. 

\begin{figure}
\caption{Formulation object (\keyw{wirepair.pro}, part 5 of 7).\label{figPart5}}
\begin{lstlisting}[linerange={112-144}]
Formulation {
 { Name FrmVn; Type FemEquation; // V nonconductor
   Quantity {
    { Name vn; Type Local; NameOfSpace HgradVn; }
   }
   Equation {// Electrostatics (Dirichlet source)
    Integral { [ eps[] * Dof{d vn}, {d vn} ]; 
     Integration I1; Jacobian J1; In VolNoCond; }
   }
 } 
 { Name FrmVc; Type FemEquation; // V conductor
   Quantity {
    { Name vc; Type Local; NameOfSpace HgradVc; }
   }
   Equation {// Electrokinetics (Dirichlet source)
    Integral { [ sigma * Dof{d vc}, {d vc} ]; 
     Integration I1; Jacobian J1; In VolCond; }
   }
 } 
 { Name FrmA; Type FemEquation; // A all
   Quantity {// vn not used, for post-process only
    { Name vn; Type Local; NameOfSpace HgradVn; }
    { Name vc; Type Local; NameOfSpace HgradVc; }
    { Name a;  Type Local; NameOfSpace HcurlA; }
   }
   Equation {// Magnetostatics (J=sigma*E source)
    Integral { [ 1 / mu[] * Dof{d a}, {d a} ];
     Integration I1; Jacobian J1; In VolAll; }
    Integral { [ sigma * {d vc}, {a} ];
     Integration I1; Jacobian J1; In VolCond; }
   }
 }
}
\end{lstlisting}
\end{figure} 

Equations~(\ref{eqfrmEsm}-\ref{eqfrmBsm}) appear as the \keyw{FrmVn}, \keyw{FrmVc} and \keyw{FrmA} formulations in Fig.~\ref{figPart5}. 
All \keyw{Equation} expressions are implicitly equal to zero. 
The \keyw{Dof} keyword (Degrees of freedom) is not used in the second integral in \keyw{FrmA} since the electric potential in~$\Omega_c$ is precalculated by the \keyw{FrmVc} formulation and is known when \keyw{FrmA} is solved.

\subsection{Resolution}
\label{ssec_res}

The Resolution object in Fig.~\ref{figPart6} triggers the actual computation for the three formulations described previously. 
The discretized equations are expressed in matrix form (\keyw{Generate} keyword), then solved iteratively using the method of conjugate gradients~\cite{JIN2014}. 
Since the matrices do not explicitly appear in the GetDP syntax, they are not considered here. 

\begin{figure}
\caption{Resolution object (\keyw{wirepair.pro}, part 6 of 7).\label{figPart6}}
\begin{lstlisting}[firstline=146, lastline=159]
Resolution {
 { Name ResAll;
   System {
    { Name SVn; NameOfFormulation FrmVn; }
    { Name SVc; NameOfFormulation FrmVc; }
    { Name SA;  NameOfFormulation FrmA; }
   }
   Operation {
    Generate[SVn]; Solve[SVn]; SaveSolution[SVn];
    Generate[SVc]; Solve[SVc]; SaveSolution[SVc];
    Generate[SA];  Solve[SA];  SaveSolution[SA];
   }
 }
} 
\end{lstlisting}
\end{figure} 

The potential associated with \keyw{FrmVc} must be solved before  \keyw{FrmA}, such that it can be used as the source term in \keyw{FrmA}. 
More advanced Resolution commands provide a powerful means to step multiple formulations in tandem, suitable for solving multiphysics problems. An example of this is a thermal-electric problem, where the resistance depends on the temperature, and the temperature in turn depends on the dissipated power (thus the resistance).

\subsection{PostProcessing and PostOperation}
\label{ssec_post}
Once the problem is solved, two final objects allow the computation of auxiliary fields or post-processing integrals. Because only one fourth of the domain is simulated, the integrals in this section are multiplied by four to obtain the true quantities for the wirepair.

The power and energy calculations in eqs.~(\ref{eqpc}) and (\ref{eqwbwe}) appear as the integrals in the top half of Fig.~\ref{figPart7}. 
For the simulation, these integrals correspond to
\begin{equation}
\label{eqpc4}
P_c=4\int_{\Omega_c}\sigma \|\vect{E}_c\|^2\,d\Omega_c,
\end{equation}
\begin{equation}
\label{eqwb4}
W_B=4\int_{\Omega}\frac{1}{2\mu} \|\vect{B}\|^2\,d\Omega,
\quad~~
W_E=4\int_{\Omega_n}\frac{\epsilon}{2} \|\vect{E}_n\|^2\,d\Omega_n.
\end{equation}

\begin{figure}
\caption{PostProcessing and PostOperation objects (\keyw{wirepair.pro}, part 7 of 7).\label{figPart7}}
\begin{lstlisting}[linerange={161-175,180-190,193-200,204-206}]
PostProcessing {
 { Name PostRLC; NameOfFormulation FrmA;
   Quantity {
    { Name Pc; Value {Integral {Type Global; 
      [ 4 * sigma * SquNorm[-{d vc}] ];
      Integration I1; Jacobian J1; In VolCond; }}
    }
    { Name Wb; Value {Integral {Type Global; 
      [ 4 / (2*mu[]) * SquNorm[{d a}] ];
      Integration I1; Jacobian J1; In VolAll; }}
    }
    { Name We; Value {Integral {Type Global; 
      [ 4 * (eps[]/2) * SquNorm[-{d vn}] ];
      Integration I1; Jacobian J1; In VolNoCond;}}
    }
   }
 }
}

PostOperation {
 { Name PostRLC; NameOfPostProcessing PostRLC; 
   Format Table;
   Operation {
    Print[ Pc, OnGlobal, StoreInVariable $Pc ];
    Print[ Wb, OnGlobal, StoreInVariable $Wb ];     
    Print[ We, OnGlobal, StoreInVariable $We ];
    Print[ {$I = $Pc / (2*(vIn-vOut))}, Format 
     "I = %.10g [A]", File > "output.txt" ];
    Print[ {2*(vIn-vOut) / ($I*ell) *10^3}, Format 
     "R = %.10g [mOhm/m]", File > "output.txt" ];
    Print[ {2*$Wb / ($I^2*ell) *10^9}, Format 
     "L = %.10g [nH/m]", File > "output.txt" ];
    Print[ {2*$We / (4*vIn^2*ell) *10^12}, Format 
     "C = %.10g [pF/m]", File > "output.txt" ];
   }
 } 
}
\end{lstlisting}
\end{figure}

The $\vect{E}$ and $\vect{B}$ fields are calculated with the gradient and curl of their respective potentials, as shown in eqs.~(\ref{eqmaxpot}) and~(\ref{eqJE}), using the exterior derivative operator \keyw{d}. 
The integral results are stored in variables for computing the final post-processing quantities, as is visible in the PostOperation object (bottom half of Fig.~\ref{figPart7}).

\section{Simulation Results}
\label{sec_res}

The simulation results for uninsulated wires can be compared with known analytical RLC expressions to help validate the simulation, after which the effects of the insulation and the defect can be studied. The results are also verified with the Altair Flux software, using a fourth order IABC open boundary. All numerical values given are obtained with 3D simulations.

The low-frequency resistance, capacitance, and inductance of the uninsulated wirepair ($\epsilon_i=\epsilon_0$), per unit length, are~\cite{NEFF1991}
\begin{equation}
\label{eqrpcp}
R=\frac{2\rho}{\pi (r_c)^2},\qquad C=\frac{\pi\epsilon_0}{\cosh^{-1}(a/r_c)},
\end{equation}
\begin{equation}
\label{eqlp}
L=\frac{\mu_0}{\pi}\left(\ln\left(\frac{2a}{r_c}\right)+\frac{1}{4}\right).
\end{equation}

Results for the above quantities appear in Table~\ref{tabValid}, for both the Altair Flux and ONELAB (Gmsh and GetDP) applications. 
The relative errors in the table are obtained by evaluating the difference between the simulation and analytical quantities and then dividing by the analytical quantity. They are expressed in parts per million (ppm).

\begin{table}
\caption{Simulation results for uninsulated wires.\label{tabValid}}
\vspace{2pt}
\begin{tabular}{c c c c}
 \noalign{\hrule height 1pt}			
			& $R$				& $L$ 				& $C$ 				\\
			& (m$\Omega$/m) 	& (nH/m)			& (pF/m) 			\\\noalign{\hrule height 1pt}	
~~Analytical~~	& ~~$104.3800020$~~ 	& ~~$732.1801501$~~	& ~~$18.12037190$~~	\\\hline
Altair Flux	& $104.3800493$	& $732.1763450$ 	& $18.12046880$ 	\\ 
(Rel. err.)	& ($0.453$~ppm)	& ($-5.20$~ppm)	& ($5.35$~ppm)		\\\hline	
ONELAB		& $104.3800311$	& $732.1782700$ 	& $18.12051767$ 	\\ 
(Rel. err.)	& ($0.279$~ppm)	& ($-2.57$~ppm)	& ($8.04$~ppm)		\\\noalign{\hrule height 1pt}
\end{tabular}
\end{table}

\begin{figure}
\caption{Electric field in the nonconducting region (top) and magnetic field (bottom) for the uninsulated case (2D). \label{figFields}}
\vspace{5pt}
\centerline{\includegraphics[width=3.35in]{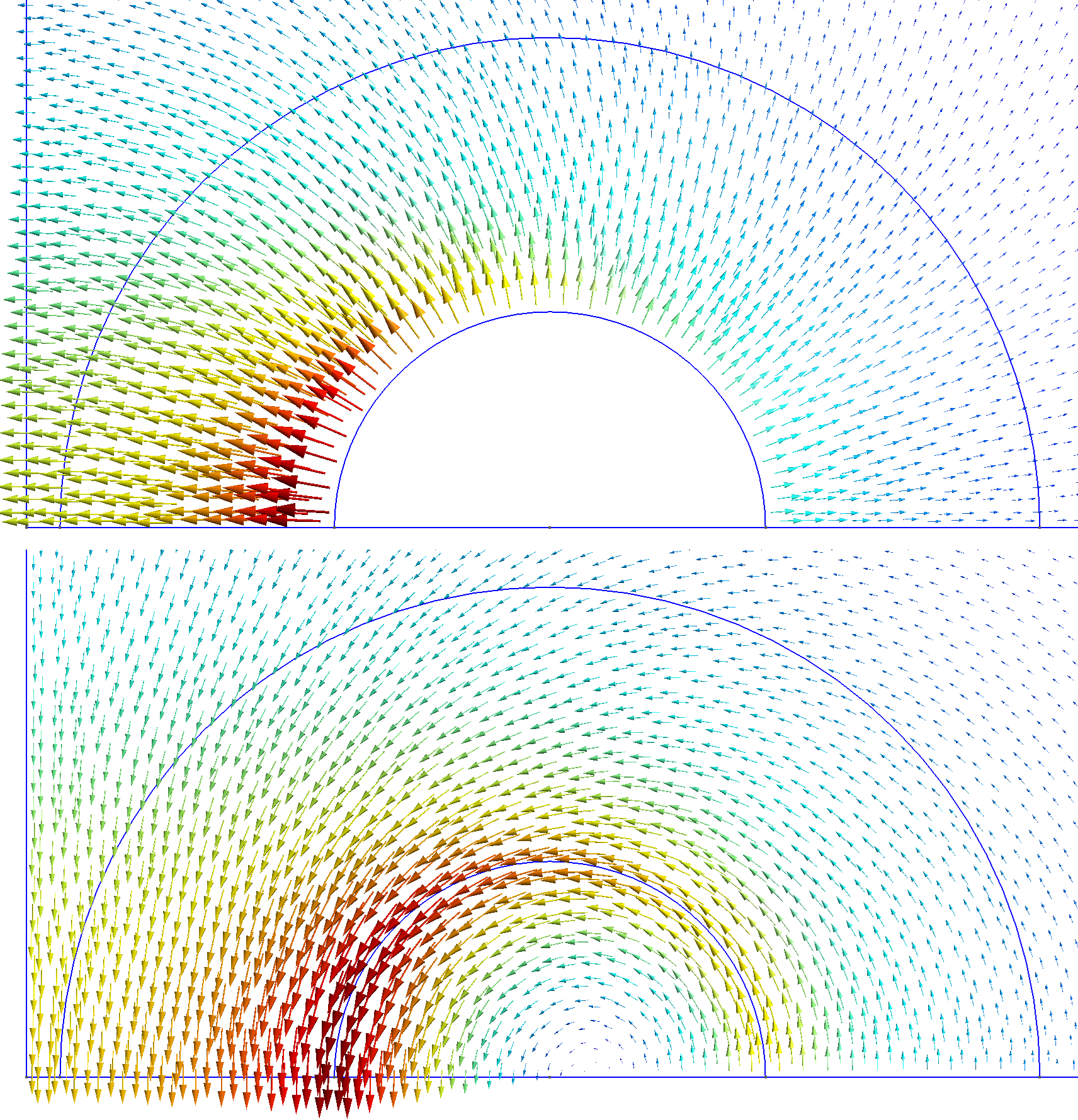}}
\end{figure}

The electric and magnetic fields are shown in Fig.~\ref{figFields}. Maximum field values are 32~kV/m and 0.375~mT, which occur at the leftmost point of the conductor. The electric field points outwards of the conductor, and the magnetic field is counter clockwise, as expected for the right wire (wire~2). The current in the wires is 0.4790188264~A. 

After adding the insulators (by setting $\epsilon_i=2.25\epsilon_0$), the capacitance becomes $28.71658404$~pF/m, an increase of 58.5\% compared to the bare wires. This also represents a relative difference of $3.36$~ppm when compared to the result obtained with Altair Flux ($28.71648756$~pF/m).

\begin{table}
\caption{Simulation results for insulated wires with the defect shown in Fig.~\ref{figCut}.\label{tabExpRes}}
\vspace{2pt}
\begin{tabular}{c c c c}
\noalign{\hrule height 1pt}				
				& $R$				& $L$ 				& $C$ 				\\
				& (m$\Omega$/m) 	& (nH/m)			& (pF/m) 			\\\noalign{\hrule height 1pt}
~~Altair Flux~~		& ~~$107.4599143$~~ 	& ~~$736.3836210$~~	& ~~$27.64998724$~~	\\
ONELAB			& $107.4590667$	& $736.3354386$ 	& $27.65036262$	\\\hline
(Rel. diff.) 		& ($-7.89$~ppm)	& ($-65.4$~ppm)	& ($13.6$~ppm)		\\\noalign{\hrule height 1pt}
\end{tabular}
\end{table}

The next simulation consists in evaluating the presence of a defect in the insulated wires (shown in Fig.~\ref{figCut}). Table~\ref{tabExpRes} shows that the defect increases the resistance and the inductance, but decreases the capacitance since a small portion of the insulation is removed. 

When removing the entire insulation, but keeping the defect in the metal, the capacitance decreases slightly to $18.09625476$~pF/m ($-0.134$\%) when compared to the undamaged bare metal. ONELAB simulation results obtained with the IABC open boundary method are similar to those reported above, but are omitted for brevity; these simulation files are available in the supplementary material~\cite{SUPMAT}.

\begin{figure}
\caption{Mesh refinement for the ONELAB $C$ result of Table~\ref{tabValid}.\label{figConv}}
\vspace{4pt}
\centerline{\includegraphics[width=3.15in]{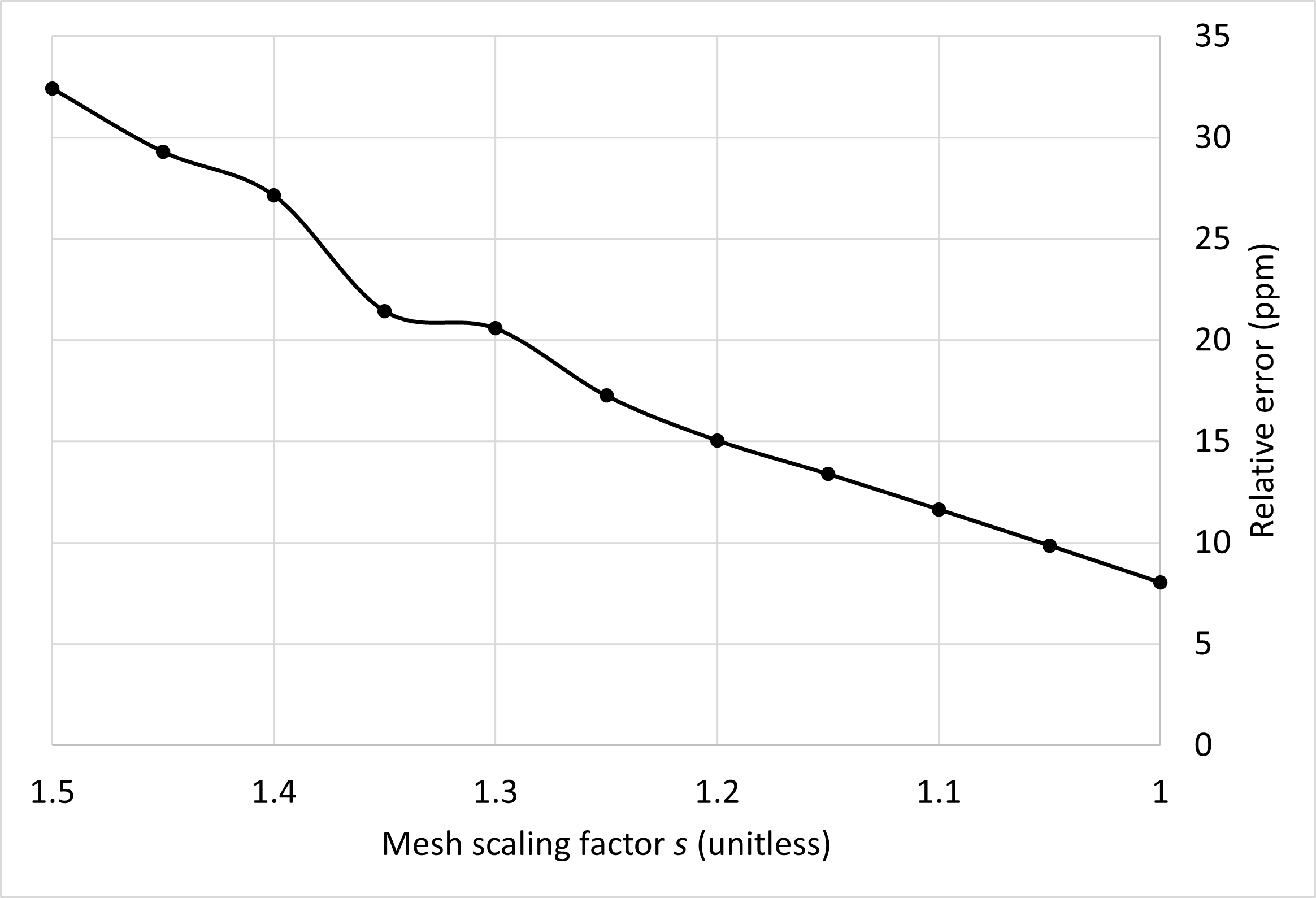}}
\end{figure}

In order to provide an appreciation for the accuracy that can be expected by the simulation, a mesh refinement graph can be devised. This graph shows the relative error as the mesh is progressively refined, such that convergence can be evaluated. Fig.~\ref{figConv} shows the relative error for the ONELAB capacitance simulation of Table~\ref{tabValid} as the mesh is progressively refined. 
The horizontal axis spans from a coarse mesh ($s=1.5$) to a fine mesh ($s=1$), from left to right, where $s$ is the mesh scaling parameter shown near the end of the \keyw{.geo} file (Appendix~A).
The relative error decreases as the mesh gets finer, and from this graph, it can be expected that continuing to refine the mesh would lead to an even lower relative error. When a relative error cannot be calculated (i.e., when no analytical reference value exists), then the convergence can be assessed by directly graphing the post-processing result itself ($R$, $L$ or $C$, for example).

Experimental note: The largest ONELAB simulation comprises approximately 1.4M unknowns, with a mesh of roughly 320k nodes and 235k elements; on a Windows system with an Intel Core i7-8700 processor (12 logical processors, 3.2 GHz), with 32~GB of RAM, the memory usage and wall-clock simulation time are on the order of 27~GB and 9 minutes, or 17~GB and 3 minutes with the real-only version of GetDP (without support for complex numbers). 
Simulation runtimes and memory usage for the Altair Flux simulations are roughly comparable to those of ONELAB and are omitted. The meshing was kept as similar as possible for both tools, differing only in their exterior regions. 
The software versions used are: GetDP 3.5.0, Gmsh 4.14.1, Altair Flux 2025.1.1.

\section{Conclusion}
\label{sec_concl}

The finite element method is an important computational tool used to solve many physics and engineering problems involving partial differential equations. 
The DC (or low-frequency) RLC parameters of a pair of parallel wires were computed with 3D simulations, necessary to support the proposed defect. Detailed steps showed how to model an open boundary problem using an open-source software package. 
Simulation results compare favourably to those obtained using analytical calculations and the Altair Flux solver. 

A brief overview of many finite element notions such as boundary conditions and weak formulations were also covered. A 3D modeling of electrostatics, magnetostatics and electrokinetics was given, which can then be used to model many other static-field configurations involving conductors and dielectrics. This tutorial provides a solid foundation upon which to explore spherical or box transformations, along with 2D planar or axisymmetrical problems. Other topics beyond the scope of this tutorial include the modeling of permanent magnets, force calculations and high-frequency applications.

Although the simulations were performed using a specific FEM software package, most of the concepts apply to other simulation tools. In some cases, the formulations and/or basis functions are hidden behind the software's front-end; however, knowledge of these concepts is very useful when diving deeper into the finite element method.

\begin{acknowledgments}
The author would like to thank Danielle Bouthot, Geneviève Savard, Guillaume Roy-Fortin, Alain Hénault and the anonymous reviewers for many suggestions and discussions that helped improve the paper.
\end{acknowledgments}

\def\bibsection{\section*{References}}

\appendix
\renewcommand\thefigure{\thesection\arabic{figure}}

\section{Geometry Specification}
\label{apx_geo}
\setcounter{figure}{0}

The geometrical model's code listing (without the defect) is shown in Fig.~\ref{figGeo}. 
The first declarations are the geometrical elements (\keyw{Cylinder}), along with the \keyw{BooleanFragments} operator, which partitions the intersecting volumes and surfaces into distinct regions. The physical volumes and surfaces are then declared, based on numbers obtained by analyzing the different Elementary Entities in the Visibility dialog in Gmsh. 

\begin{figure}[b]
\caption{3D model (\keyw{wirepair.geo}). The version with the defect (``V''-grove) is not shown for reasons of space, but is available online~\cite{SUPMAT}.\label{figGeo}}
\begin{lstlisting}[linerange={2-3,13-27,94-96,100-103,115}]
Include "wirepaircommon.pro";
SetFactory("OpenCASCADE");
Cylinder(1) = {a,0,0, 0,0,ell, rc,Pi};  // cond.
Cylinder(2) = {a,0,0, 0,0,ell, ri,Pi};  // insul.
Cylinder(3) = {0,0,0, 0,0,ell, rb,Pi/2};// bnd int
Cylinder(4) = {0,0,0, 0,0,ell, re,Pi/2};// bnd ext
BooleanFragments{ Volume{4}; Delete; }{ 
 Volume{1:3}; Delete; }

Physical Volume("Wire cond", 100) = {1};
Physical Volume("Wire insul", 101) = {4};
Physical Volume("Vacuum int", 102) = {3};
Physical Volume("Vacuum ext", 103) = {2}; // shell
Physical Surface("Left", 120) = {3, 11};
Physical Surface("Front", 121) = {2, 7, 13, 18};
Physical Surface("Back", 122) = {4, 12, 17, 19};
Physical Surface("Ext", 123) = {1}; // curved r=re

// Mesh parameters
s = 1.0;// scaling factor: 1.0=fine, 1.5=coarse
MeshSize{:}                    = s*0.08*mm;
MeshSize{PointsOf{Volume{1};}} = s*0.04*mm;//cond.
MeshSize{PointsOf{Volume{2};}} = s*0.12*mm;//shell
MeshSize{PointsOf{Surface{1};}}= s*0.20*mm;//ext
Mesh.HighOrderOptimize = 2; Mesh.ElementOrder = 2;  

Color Black{ Surface{:};}
\end{lstlisting}
\end{figure}

The last section of the geometry file concerns meshing parameters. In the present case, a denser mesh near the conductor improves the accuracy of the simulation, since this is where the fields have the highest magnitudes and the greatest variations (visible in Fig.~\ref{figFields}). As expected, mesh density, simulation accuracy and computation time/memory are interrelated.


\begin{thebibliography}{99}

\bibitem{NEFF1991} H. P. Neff, \textit{Introductory Electromagnetics}, (John Wiley \& Sons, New York, NY, 1991), p.~300.

\bibitem{HAZIM2022} K. Hazim, G. Parent, S. Duchesne, A. Nicolet and C. Geuzaine, ``2D electrostatic modeling of twisted pairs", \textit{Intl. J. for Computation and Mathematics in Electrical and Electronic Engineering} \textbf{41}(1), 48-63 (2022).

\bibitem{MEEKER2020} D.~C. Meeker, ``User's Manual: Finite Element Method Magnetics, Version 4.2", \url{https://www.femm.info}.

\bibitem{KARBAN2025} P. Karban, D. P\'anek and J. Kaska, ``Open-source platform for simulation of physical fields: Agros", \textit{Journal of Computational and Applied Mathematics} \textbf{465}, 116589 (2025).

\bibitem{BOULE2014}
M. Boul\'e, ``The role of finite element method software in the teaching of electromagnetics", \textit{Fourth Interdisciplinary Engineering Design Education Conference}, 44-51 (2014).

\bibitem{ELMERWEB} \url{https://www.elmerfem.org}

\bibitem{FREEFEMWEB} \url{https://freefem.org}

\bibitem{NGSOLVEWEB} \url{https://ngsolve.org}

\bibitem{ONELABWEB} \url{https://onelab.info}

\bibitem{ALTAIRWEB} \url{https://altair.com/flux}

\bibitem{ANSYSWEB} \url{https://www.ansys.com/products/electronics/ansys-maxwell}

\bibitem{COMSOLWEB} \url{https://www.comsol.com}

\bibitem{IOAN2021} D. Ioan, G. Ciuprina and W. Schilders, ``Chapter 5: Complexity reduction of electromagnetic systems'', in \textit{Model Order Reduction, Volume 3: Applications}, edited by P. Benner, S. Grivet-Talocia, A. Quarteroni, G. Rozza, W. Schilders and L. Miguel Silveira (De Gruyter, Berlin, Boston, 2021), pp. 145-200.

\bibitem{GMSH2024} C. Geuzaine and J.-F. Remacle, ``Gmsh Reference Manual: the documentation for Gmsh, a finite element mesh generator with built-in pre- and post-processing facilities", \url{https://gmsh.info}.

\bibitem{GETDP2024} P. Dular and C. Geuzaine, ``GetDP Reference Manual: the documentation for GetDP, a general environment for the treatment of discrete problems", \url{https://getdp.info}.

\bibitem{SUPMAT} Simulation source code files available at \url{https://github.com/MarcBoule/OnelabModels/tree/main/WirepairRLC}.

\bibitem{PAUL2008} C.~R. Paul, \textit{Analysis of Multiconductor Transmission Lines}, 2nd edition (Wiley-Interscience, Hoboken, NJ, 2008), Sec.~4.2.1.1.

\bibitem{DULAR1998} P. Dular, W. Legros and A. Nicolet, ``Coupling of local and global quantities in various finite element formulations and its application to electrostatics, magnetostatics and magnetodynamics", \textit{IEEE Transactions on Magnetics} \textbf{34}(5), 3078-3081 (1998).

\bibitem{PIRIOU2024} F. Piriou and S. Clénet, \textit{Finite Element Method to Model Electromagnetic Systems in Low Frequency}, (John Wiley \& Sons, Inc., Hoboken, NJ, 2024).

\bibitem{ZHU2006} Y. Zhu and A. C. Cangellaris, \textit{Multigrid Finite Element Methods for Electromagnetic Field Modeling}, 5th ed. (Wiley-IEEE Press, Piscataway, NJ, 2006), Secs. 2.5, 3.1, 3.2.

\bibitem{CHEN1997} Q. Chen and A. Konrad, ``A review of finite element open boundary techniques for static and quasi-static electromagnetic field problems", \textit{IEEE Transactions on Magnetics} \textbf{33}(1), 663-676 (1997).

\bibitem{NABIZADEH2021} M.~S. Nabizadeh, R. Ramamoorthi and A. Chern, ``Kelvin transformations for simulations on infinite domains", \textit{ACM Transactions on Graphics} \textbf{40}(4), Article 97 (2021).

\bibitem{REMACLE1994} J.F. Remacle, A. Nicolet, A. Genon and W. Legros, ``Comparison of boundary elements and transformed finite elements for open magnetic problems", \textit{WIT Transactions on Modelling and Simulation} \textbf{7}, 8 pages (1994).

\bibitem{HENROTTE1999} F. Henrotte, B. Meys, H. Hedia, P. Dular and W. Legros, ``Finite element modelling with transformation techniques", \textit{IEEE Transactions on Magnetics} \textbf{35}(3), 1434-1437 (1999).

\bibitem{MEEKER2014} D. Meeker, ``Improvised asymptotic boundary conditions for electrostatic finite elements", \textit{IEEE Transactions on Magnetics} \textbf{50}(6), 7400609 (2014).

\bibitem{MEEKER2013} D. Meeker, ``Improvised open boundary conditions for magnetic finite
elements'', \textit{IEEE Transactions on Magnetics}, \textbf{49}(10), 5243-5247 (2013).

\bibitem{JIN2014} J.-M. Jin, \textit{The Finite Element Method in Electromagnetics}, 3rd ed. (John Wiley \& Sons Inc., Hoboken, NJ, 2014), Secs. 5.3.4, 5.7.4, 6.3, 14.3.1, 15.2.

\bibitem{BASTOS2014} J.~P.~A. Bastos and N. Sadowski, \textit{Magnetic Materials and 3D Finite Element Modeling} (Taylor \& Francis Group, Boca Raton, FL, 2014), Chap.~5.

\bibitem{BIRO1998} O. Biro, K. Preis and C. Paul, ``The use of a reduced vector potential Ar formulation for the calculation of iron induced field errors", \textit{Proc. 1st Int. ROXIE Users Meeting Workshop.}, 31-46 (1998).

\bibitem{CREUSE2019} E. Creus\'e, P. Dular and S. Nicaise ``About the gauge conditions arising in finite element magnetostatic problems", \textit{Computers \& Mathematics with Applications} \textbf{77}(6), 1563-1582 (2019).

\bibitem{GEUZAINE2001} C. Geuzaine, \textit{High Order Hybrid Finite Element Schemes for Maxwell’s Equations Taking Thin Structures and Global Quantities into Account}, Ph.D. Thesis, University of Li\`ege, Belgium, 2001.

\bibitem{DHATT2012} G. Dhatt, G. Touzot and E. Lefran\c{c}ois, \textit{Finite Element Method} (ISTE Ltd and John Wiley \& Sons, London UK and Hoboken NJ, 2012), Secs. 2.5, 5.1.

\end{thebibliography}
\end{document}